\documentstyle[aps,prl,epsfig,floats]{revtex}
\begin{document}
\draft
\twocolumn[
\hsize\textwidth\columnwidth\hsize\csname@twocolumnfalse\endcsname
\title{Chiral hedgehog textures in 2D $XY$-like ordered domains}
\author{Kok-Kiong Loh}
\address{Department of Physics, University of California
at Los Angeles, California 90095-1547}
\author{Isabelle Kraus}
\address{Institut de Physique et Chimie des Mat\'{e}riaux,
Groupe des Mat\'{e}riaux Organiques, Unit\'{e} Mixte
CNRS-Universit\'{e} Louis Pasteur, 23 rue du Loess, F-67037
Strasbourg Cedex, France}
\author{Robert B. Meyer}
\address{The Martin Fisher School of Physics, Brandeis University,
Waltham, Massachusetts 02254-9110}
\date{\today}
\maketitle
\begin{abstract}
The textures associated with  a point defect centered in a
circular domain of a thin film with $XY$-like ordering have been
analyzed.  The family of equilibrium textures, both stable and
metastable, can be classified by a new radial topological number
in addition to the winding number of the defect.  Chiral textures
are supported in an achiral system as a result of spontaneously
broken chiral symmetry. Among these chiral textures, our
theoretical analysis accurately describes two categories of
recently discovered ``reversing spiral'' textures, ones that are
energetically stable and metastable.
\end{abstract}
\pacs{68.10.Cr, 68.18.+p, 68.55.Ln, 68.60.-p} ] Thin films of
elongated molecules with tilt ordering, including Smectic~C liquid
crystals and dense fluid phases of amphiphiles deposited on water,
very often possess fascinating distributions of the tilt azimuth.
The organization of the tilt azimuth is referred to as the texture
and can be observed by polarized light microscopy or Brewster
angle microscopy. Many classes of these textures have been found
experimentally, like stripes\cite{stripe}, stars\cite{star},
boojums\cite{boojums} and hedgehogs\cite{deMul}. These examples
are observed in Langmuir monolayers composed of molecules which
are symmetric under in-plane reflection, or achiral. Recently,
more attention has been paid to the hedgehog patterns, obtained in
a circular domain with a central point defect. The reversing
spiral is one of the spectacular hedgehog textures discovered in a
chiral tilted Smectic~C liquid crystal film on water~\cite{Kraus}.
Let us emphasize that both systems described above are polar, they
are not symmetric under 180$^\circ$ rotation about an in-plane
axis. The tilt azimuth can be represented by a two-dimensional
vector $\hat{c}$, the projection onto the film of the elongated
molecules, analogous to the order parameter of an $XY$-model.
Textures in two-dimensional $XY$-like systems have attracted a
good deal of attention. A brief theoretical account based on a
perturbative approach of hedgehogs in achiral monolayers can be
found  in Ref.~\cite{FiscBru}. The stability of the hedgehog
configuration in an isotropic and achiral system has also been
investigated~\cite{PettyLuben}. Spiral textures in an achiral
system have been studied in the small distortion regime
\cite{Williams}. Although a theoretical approach to solutions for
hedgehog textures is given in Ref.~\cite{Kraus}, to the best of
our knowledge, a systematic discussion of hedgehog textures in a
chiral system has not been presented.

In this work,  we study a generic model \cite{LangSeth} for the
texture in a circular domain of a two-dimensional $XY$-like
system. We have found a family of equilibrium configurations that
can be classified by a new topological number analogous to the
winding number that classifies a two-dimensional point
defect\cite{Mermin}.  When a system is achiral, these metastable
configurations can be chiral as a result of spontaneously broken
chiral symmetry. When chirality is explicitly introduced,  more
complex equilibrium textures result.  The crossed polarizer images
generated from the theoretical textures are in excellent agreement
with the pictures taken experimentally. The starting point of our
investigation is the following elastic energy of a chiral polar
film with tilt ordering,
\begin{eqnarray}
H[\hat{c}]&=&\frac{1}{2}\int_\Omega dA \left[K_s|\nabla\cdot\hat{c}|^2
+K_{sb}(\nabla\cdot\hat{c})(\nabla\times\hat{c}\cdot\hat{z})+
\right.\nonumber\\
&& \left.K_b|\nabla\times\hat{c}\cdot\hat{z}|^2\right]+
\oint_\Gamma ds\;\sigma(\vartheta-\Theta).
\end{eqnarray}
It is computed for a circular area $\Omega$ containing the ordered
medium enclosed by the boundary $\Gamma$. The unit vector
$\hat{z}$ points normal to the film. The quantities $K_b$ and
$K_s$ are, respectively, the bend and splay elastic moduli,
$\hat{c}\equiv\hat{x}\cos\Theta+\hat{y}\sin\Theta$ is the order
parameter of the system, $\Theta$ is the angle between $\hat{c}$
and the $x$ axis, $\vartheta$ is the angle between the outward
normal to $\Gamma$ and the $x$ axis. The anisotropic line tension
$\sigma(\phi)$ can formally be expanded as
$\sigma_0+\sum_n(a_n\cos n\phi+c_n \sin n\phi)$. We will consider
only the first few coefficients in the expansion. The first order
terms are $a_1$ for polar films and $c_1$ for chiral systems.
When the system is nonpolar, $a_1$ vanishes and $a_2$ must be
considered. The coefficient $c_2$ is relevant when the system
is both polar and chiral, but it will be neglected because
$a_1$ and $c_1$ are non zero for such a system. The cross
term$(\nabla\cdot\hat{c})(\nabla\times\hat{c}\cdot\hat{z})$ has to
be included when the film is both chiral and polar. It is required
that the coefficient $|K_{sb}|<2\sqrt{K_sK_b}$ so that the elastic
energy density remains positive for arbitrary splay and bend
distortions.

We will restrict ourselves to the hedgehog textures, each
containing a central point defect of winding number $+1$ with core
radius $\xi$.  The defect core corresponds to the region in which
$\hat{c}$ is not defined. We assume that its presence affects the
elastic energy  only through the inner boundary condition at
$r=\xi$ and we neglect its energetic contribution when $r<\xi$
\cite{deGennes}. The boundary condition is taken to be
$\Theta|_{r=\xi}=\varphi+\varphi_d$, where $r$ is the radial
distance, $\varphi$ is the polar angle in plane-polar coordinates
and $\varphi_d$ is a constant. This is indeed justifiable for
$\pm1$ defects with the structure discussed in Ref.~\cite{Tabe}.
We further assume that the defect is stable when it is located at
the center of $\Omega$ and that the system is cylindrically
symmetric, i.e., $\Theta=\varphi+f(\ln r)$, where $f(\ln r)$ is
the radial distribution of $\hat{c}$. These assumptions can be
shown to be valid for all the textures to be discussed. We shall
compute the possible expressions for $f(\ln r)$ that minimize the
elastic energy and give the equilibrium textures. In terms of
$k\equiv\ln r$, the elastic energy reduces to
\begin{eqnarray}
H[f, f^\prime]&=&\pi\kappa\int_{\ln\xi}^{\ln R_0}dk\left\{1+f^{\prime2}-
\right.\nonumber\\
&&\mu\left[(1-f^{\prime2})\cos(2f-2\varsigma_+)-\right.\nonumber\\
&&\left.\left.2f^\prime\sin(2f-2\varsigma_+)\right]
\right\}+\left.2\pi R_0\sigma(-f)\right|_{\ln R_0},\label{reducedH}
\end{eqnarray}
where $2\kappa\equiv K_s+K_b$, $2\kappa\beta\equiv K_s-K_b$,
$2\kappa\tau\equiv K_{sb}$, $\mu\equiv\sqrt{\beta^2+\tau^2}$ and
$2\varsigma_\pm\equiv\tan^{-1}\tau/\beta\pm\pi$. The equilibrium
condition for $f(k)$ is
\begin{eqnarray}
-f^{\prime\prime}-\mu\left[f^{\prime\prime}\cos(2f-2\varsigma_+)-
\right.&&\nonumber\\
\left.(f^{\prime2}+1)\sin(2f-2\varsigma_+)\right]&=&0\label{bulkr}
\end{eqnarray}
and the boundary condition at $k=\ln R_0$ is
\begin{eqnarray}
\kappa\left\{f^\prime+\mu\left[f^\prime\cos(2f-2\varsigma_+)+\sin(2f-2
\varsigma_+)\right]\right\}-&&\nonumber\\
R_0\sigma^\prime(-f)&=&0.
\end{eqnarray}
Depending on the choice of the parameters, $f$ may possess more
than one or, at times, numerous solutions. We use linear stability
analysis to determine if these solutions are local minima.  The
elastic energy is expanded to second order in small variations
$\chi\equiv f-f_0$ about an equilibrium configuration $f_0$ as
$\delta H=\int dk\chi{\cal L}\chi + {\psi}^T{\cal B}{\psi}$ where
\begin{eqnarray}
{\cal L}&=&-\pi\kappa\left\{\left[1+\mu\cos(2f_0-2\varsigma_+)
\right]\frac{d^2}{dk^2}-\right.\nonumber\\
&&\left.\mu\left[f^{\prime\prime}_0\sin(2f_0-2\varsigma_+)+
2\cos(2f_0-2\varsigma_+)\right]
\right\},\\
{\cal B}_{\chi^\prime\chi^\prime}&=&0,\\
{\cal B}_{\chi\chi^\prime}&=&{\cal B}_{\chi^\prime\chi}=\frac{\pi\kappa}{2}
\left[1+\mu\cos(2f_0-2\varsigma_+)\right]_{\ln R_0},\\
{\cal B}_{\chi\chi}&=&\pi\left\{-\kappa\mu\left[f^\prime_0\sin(2f_0-2\varsigma_+)-2
\cos(2f_0-2\varsigma_+)\right]+\right.\nonumber\\
&&\left.R_0\sigma^{\prime\prime}(-f_0)\right\}_{\ln R_0},
\end{eqnarray}
and $\psi^T\equiv(\chi^\prime,\chi)|_{\ln R_0}$.
The deviation of the elastic energy $\delta H$ from its
equilibrium value can be examined in terms of the eigenvalue
$\lambda$ and associated eigenfunction $\phi_\lambda$ satisfying
${\cal L} \phi_\lambda=\lambda\phi_\lambda$. The eigenfunctions
$\phi_\lambda$ are normalized so that $\int dk \phi_\lambda^2=1$
and $\phi_\lambda|_{\ln\xi}=0$ for infinitely strong anchoring at
the inner boundary.  It is typical in linear stability analysis
that there is another boundary condition on $\phi_\lambda$ that
isolates a set of eigenvalues $\lambda$ in which the sign of the
lowest one is to be tested. For our particular case, there is
no other restriction on $\phi_\lambda$ that can be imposed and all
$\lambda$'s are allowed.   The deviation of energy associated with
the fluctuational mode $\phi_\lambda$ is no longer $\lambda$ but
\begin{eqnarray}
\delta H_\lambda=\lambda+\psi^T_\lambda{\cal B}\psi_\lambda,
\end{eqnarray}
where $\psi^T_\lambda=(\phi^\prime_\lambda, \phi_\lambda)|_{\ln R_0}$.
Instability is signified by $\delta H_\lambda$ becoming negative.
The asymptotic behavior of $\delta  H_\lambda$ can be shown to be
$\delta H_\lambda\sim -|\lambda| + R_0^{(1+2\sqrt{|\lambda}|)}$
when $\lambda\rightarrow-\infty$, and $\delta H_\lambda> (1-{\cal
B}_{\chi\chi^\prime}/2{\cal B}_{\chi\chi})\lambda$ when
$\lambda\rightarrow\infty$ for $ R_0\sqrt{a_1^2+c_1^2}\gg\kappa$
and $\sigma^{\prime\prime} (-f_0)|_{k=\ln R_0} \approx
\sqrt{a_1^2+c_1^2}$. We see that $\delta H_\lambda$ is positive in
both limits and can only change sign for small $\lambda$.

\begin{figure}[tb]
\centerline{\epsfig{file=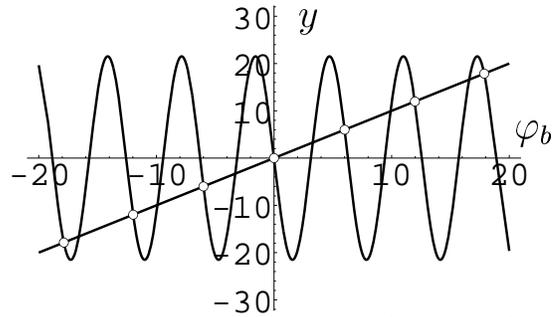, width=3in}} \caption{Plots of
$y=\varphi_b$ and $y=R_0a_1(\ln R_0-\ln\xi) \sin\varphi_b/\kappa$
versus $\varphi_b$. The intersections, $\varphi_b^{(i)}$,  give
the equilibrium configurations $f_0^{(i)}(k)$. Those marked
with open circles are stable. The parameters used are $\kappa=1$,
$R_0=5$, $\xi=0.1$, and $a_1=-1.1$.} \label{mltspr}
\end{figure}

We have outlined the procedures to obtain the equilibrium textures
and to examine their stability against infinitesimal fluctuations.
To understand these equilibrium textures,  we first look at the
simplest case $\beta=\tau=c_1=a_2=\varphi_d=0$ and $a_1<0$. It
corresponds to a polar film made of achiral molecules, with
isotropic elastic constants and having fixed anchoring at the
inner boundary such that $\hat{c}$ points normal into the bulk.
The condition $a_1<0$ indicates that $\hat{c}$ favors the outward
normal direction at the outer boundary, without being locked. In
this case, there are analytic solutions. It is obvious that
$f_0(k)=0$ is the lowest energy configuration: all $\hat{c}$
vectors point along the radial direction. The general solution to
$f_0(k)$ is
\begin{eqnarray}
f_0(k)&=&\frac{\varphi_b}{\ln R_0-\ln\xi}(k-\ln\xi),
\end{eqnarray}
and $\varphi_b$ satisfies
\begin{eqnarray}
\varphi_b&=&\frac{R_0a_1(\ln R_0-\ln\xi)}{\kappa}\sin\varphi_b.\label{bc1}
\end{eqnarray}
The quantity $\varphi_b$ is the anchoring angle of $\hat{c}$
measured with respect to $\varphi$ at $k=\ln R_0$.
It is easy to see that the system supports numerous equilibrium
solutions when the amplitude $|R_0a_1(\ln
R_0-\ln\xi)/\kappa|\gg1$.  Figure \ref{mltspr} shows the plots of
$\varphi_b$ and $R_0a_1(\ln R_0-\ln\xi)\sin\varphi_b/\kappa$. We
denote the solutions of Eq.~(\ref{bc1}) by $\varphi_b^{(i)}$ with
an index $i$.  Not all the solutions are stable.    The stable
ones are indicated by open circles in Fig.~\ref{mltspr}. When
$\varphi_b^{(i)}\neq0$,  we have spirals in which $\hat{c}$ points
in the radial direction at the inner boundary and rotates
counterclockwise through $\varphi_b^{(i)}$ along a radial path
ending at the outer boundary. These metastable textures do not
have in-plane reflection symmetry although the system in question
is achiral. We find, by inspecting the boundary condition
Eq.~(\ref{bc1}), that there is a solution
$\varphi_b^{(j)}=-\varphi_b^{(i)}$ for any solution
$\varphi_b^{(i)}$ and these configurations have the same energy.
Hence,  chiral symmetry is spontaneously broken for the higher
energy metastable configurations. The family of solutions can be
classified in terms of a topological index $m_i$, which is defined
to be the nearest integer around $\varphi_b^{(i)}/2\pi$. This
classification can also be drawn topologically by examining the
possible order-parameter distributions along a line in the radial
direction connecting the inner and the outer
boundaries\cite{Mermin}.   All integers are allowed topologically,
but only some are supported energetically in a bounded system.

\begin{figure}[tb]
\centerline{\epsfig{file=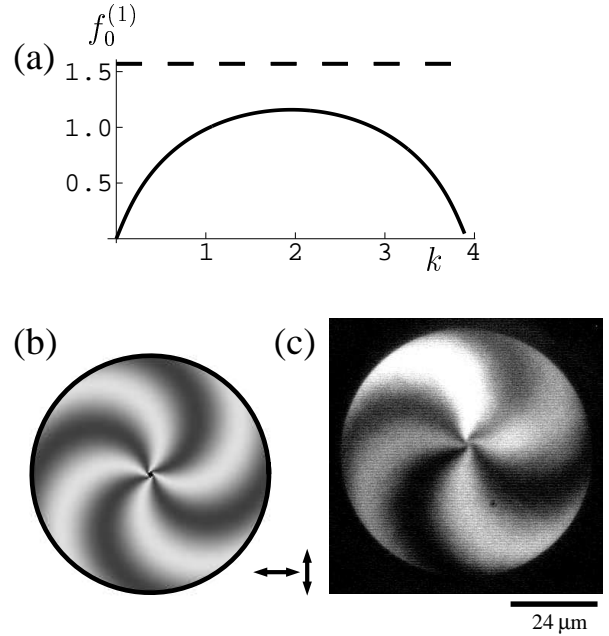, width=3.2in}} \caption
{Illustrations of the lowest energy configuration for $\kappa=1$,
$\mu=0.70$, $\varsigma_+=\pi/2$, $R_0=5$, $\xi=0.1$, $a_2=-10$,
and $\varphi_d=0$. (a) is a plot of $f_0^{(1)}(k)$ versus $k$ for
$f_0^{(1)\,\prime}(0)=1.9788$, the dashed line indicates
$\varsigma_+$, (b) shows the corresponding density plot of
$\sin^2\Theta\cos^2\Theta$, and (c) depicts the experimental image
of a reversing spiral observed in free standing film of chiral
Smectic C liquid crystal. The polarizer and the analyzer are
vertical and horizontal. } \label{lwscfg}
\end{figure}

The situation is different when $a_1>0$, i.e., when the preferred
direction of $\hat{c}$ at the outer boundary is in the inward
normal direction. The solution $f_0(k)=0$ is now unstable and we
can not have a pure splay texture anymore. The lowest energy
configuration corresponds to the first nonzero $\varphi_b^{(i)}$
satisfying Eq.~(\ref{bc1}). It is a spiral texture as are all the
other metastable configurations allowed. Let us emphasize that
they all break the chiral symmetry spontaneously. The spiral
textures we have just discussed arise from spontaneously broken
chiral symmetry in an isotropic achiral system. The actual system
may not be isotropic and the parameter $\beta$ is not necessarily
zero. Further, it is possible to explicitly break the chiral and
polar symmetries ($\tau \neq 0$) by adding chiral molecules to the
system and by putting the film on water, respectively. When
$\beta$ or $\tau$ are nonzero, analytic solutions for
Eq.~(\ref{bulkr}) are not obvious and we resort to numerical
methods.  More complex equilibrium textures with $m_i=0$ are
supported in addition to those with higher $m_i$'s. The behavior
of the $\hat{c}$-distribution in such cases can be understood
intuitively by noting the existence of two preferred bulk
directions for $\hat{c}$ when $\Theta-\varphi\in[-\pi, \pi)$,
namely $\Theta_{\pm}=\varphi+\varsigma_\pm$. Right at the inner
boundary of $\Omega$, $\Theta$ is set to $\varphi+ \varphi_d$.
Along a trajectory traversing radially into the bulk, $\Theta$
settles smoothly near one of $\Theta_\pm$, depending on which of
these gives the lower overall energy of the system ($\Theta_+$ for
our case). When it gets to the outer boundary, $\Theta$ gradually
sets itself at $\varphi+\varphi_b$. The $f^{(1)}_0(k)$ and texture
for the lowest energy configuration taking $\tau \neq 0$ are
displayed in Fig.~\ref{lwscfg}. The texture is depicted as the
density plot of $\sin^2\Theta\cos^2\Theta$, which simulates the
images obtained by a set of crossed polarizers.  Figure
\ref{lwscfg}(c) shows an experimental image of a chiral Smectic C
liquid crystal domain in a free standing film. It is very similar
to the image shown in Fig.~\ref{lwscfg}(b), computed using the
appropriate parameters for a chiral free standing film.

\begin{figure}[tb]
\centerline{\epsfig{file=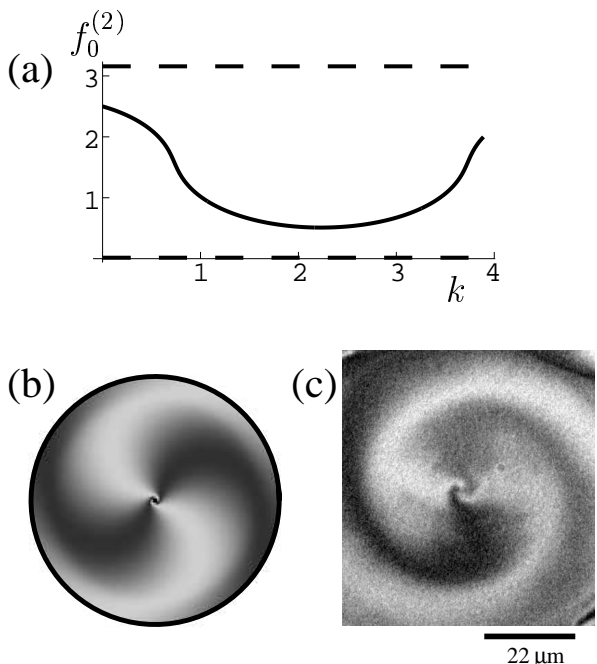, width=3.2in}} \caption
{Illustrations of a texture configuration that resembles the
reversing spiral observed experimentally in Ref.~[5], when
$f_0^{(2)\,\prime}(0)=-0.4659$, $\kappa=1$, $\beta=0.90$,
$2\varsigma_+=\pi-3.11$, $R_0=5$, $\xi=0.1$, $a_1=10$, $c_1=20$,
and $\varphi_d=2.5$. (a) shows the plot of $f^{(2)}_0(k)$ versus
$k$, the dashed lines indicate $\varsigma_+$ and
$2\pi+\varsigma_-$, (b) is the corresponding density plot of
$\cos^2\Theta$, and (c) depicts the experimental image of a DRS
observed in a chiral smectic C film on water. Polarizer and
analyzer are slightly uncrossed.} \label{revspr}
\end{figure}

Apparently, these images do not resemble the remarkable and
dramatic reversing spiral texture reported in Ref.~\cite{Kraus}
and shown in Fig.~\ref{revspr}(c). Nevertheless, Fig.~\ref{lwscfg}
does represent a reversing spiral, since $f^{(1)}_0(k)$ goes
through a maximum at about $k=2.0$.  However, the initial rotation
of the director near the core is so rapid that it is barely
visible. We are able to locate among the various solutions an
equilibrium texture that resembles the ``dramatic reversing
spiral'' (or DRS texture), illustrated in Fig.~\ref{revspr}. The
distinguishing feature of the DRS texture is the extended region
near the core in which the director rotates slowly.  This
configuration is constructed by choosing the values of $\Theta$ at
the inner boundary and at the outer boundary to be around
$\Theta_- =\varphi+\varsigma_-+2\pi$. Starting from the inner
boundary, $\Theta$ first remains near $\Theta_-$, then swings
nearly $\pi$ to a value close to $\Theta_+$ and finally returns
rapidly to $\varphi+\varphi_b$ (which has been set the be around
$\Theta_-$) near the outer boundary.  The function $f^{(2)}_0(k)$
and its texture are depicted in Fig.~\ref{revspr}. There is strong
resemblance between the density plot of $\cos^2\Theta$, simulating
the images obtained with slightly uncrossed polarizers, and the
experimental picture of the DRS.

Remarkably, the DRS is only a metastable texture!  It is easy to
see that $f^{(2)}_0(k)$ is not the lowest energy solution. The
configuration, in which $\Theta$ sets smoothly near $\Theta_-$
without going through the rapid changes to $\Theta_+$ and then
back to $\Theta_-$, has the lowest elastic energy. Since
$f^{(2)}_0(k)$ is not the global minimum of the elastic energy
Eq.~(\ref{reducedH}), its stability against fluctuations is in
question. We have examined $\delta H_\lambda$, the deviation in
energy associated with the reversing spiral solution, in a wide
range of $\lambda$, and conclude that the DRS is a metastable
configuration for the present choice of the parameters. In
general, the DRS texture is not even metastable for an arbitrary
set of parameters. As noted in Ref.~\cite{Kraus}, the DRS texture
is rare, which is consistent with these findings.

In conclusion, we have analyzed a generic model for textures
associated with a $+1$ defect in domains with $XY$-like ordering.
Restricting our analysis to the class of textures with the defect
fixed at the center, we find the equilibrium configurations and
examine the stability of these configurations against
infinitesimal fluctuations. Many metastable configurations are
supported at some suitable choice of parameters. These can be
classified by a new radial topological number analogous to the
winding number classifying the point defect. It is also found that
metastable configurations are chiral, including the stable lowest
energy texture when $a_1>0$, even if the system possesses in-plane
reflection symmetry. When elastic anisotropy, chirality and
polarity are introduced explicitly, more equilibrium solutions
exist. Among the equilibrium textures, we have found two kinds of
reversing spirals, simple ones that are absolutely stable
textures, and metastable ones of a more dramatic appearance,  that
resemble closely the reversing spiral reported in
Ref.~\cite{Kraus}.

K.-K. L. would like to express his gratitude to Professor Joseph
Rudnick for ideas, suggestions and support, and thank Professor
Charles Knobler and Professor Robijn Bruinsma for numerous
stimulating discussions. This research was supported in part by
the NSF through grant DMR-9974388, and by Brandeis University.


\begin{references}
\bibitem{stripe}
J. Ruiz-Garcia, X. Qiu, M.-W. Tsao, G. Marshall, C.M.Knobler,
G. A. Overbeck and D. M\"{o}bius,
J. Phys. Chem. {\bf 97}, 6955 (1993).
\bibitem{star}
X. Qiu, J. Ruiz-Garcia, K. J. Stine, C.M. Knobler, J.V. Selinger,
Phys. Rev. Lett. {\bf 67}, 703 (1991).
\bibitem{boojums}
S. Rivi\`{e}re and J. Meunier, Phys. Rev. Lett. {\bf 74}, 2495 (1995);
J. Fang, E. Teer, C.M. Knobler, K.-K. Loh and J. Rudnick,
Phys. Rev. E {\bf 56}, 1859 (1997).
\bibitem{deMul}  M.N.G. de Mul and J.A. Mann Jr., Langmuir {\bf 11}, 3292 (1995).
\bibitem{Kraus}
I. Kraus and R.B. Meyer, Phys. Rev. Lett. {\bf 82}, 3815 (1999).
\bibitem{FiscBru} T.M. Fischer, R.F. Bruinsma, and C.M. Knobler,
Phys. Rev. E {\bf 50}, 413 (1994).
\bibitem{PettyLuben} D. Pettey and T.C. Lubensky, Phys. Rev. E
{\bf 59}, 1834 (1999).
\bibitem{Williams}
D.R.M. Williams, Phys. Rev. E {\bf 50}, 1686 (1994).
\bibitem{LangSeth}
S.A. Langer and J.P. Sethna, Phys. Rev. A {\bf 34}, 5035 (1986).
\bibitem{Mermin} N.D. Mermin, Rev. Mod. Phys. {\bf 51}, 591 (1979).
\bibitem{deGennes} P.G. de Gennes and J. Prost, The Physics of
Liquid Crystals (Clarendon Press, Oxford, 1993), p.171.
\bibitem{Tabe}
Y. Tabe, N. Shen, E. Mazur, and H. Yokoyama,
Phys.  Rev. Lett. {\bf 82}, 759 (1999).
\end{references}
\end{document}